\documentclass[aps,showpacs,preprint,preprintnumber,nofootinbib,amsmath,amssymb,ascmac,bm,12pt]{revtex4}
\usepackage{bm}
\usepackage{graphicx}
\begin{document}
\title{
Stability of Squashed Kaluza-Klein Black Holes
}
\author{
       ${}^{1}$Masashi Kimura, 
       ${}^{2}$Keiju Murata, 
       ${}^{1}$Hideki Ishihara 
and
       ${}^{2,3}$Jiro Soda 
}
\affiliation{
${}^{1}$Department of Mathematics and Physics,
Graduate School of Science, Osaka City University,
3-3-138 Sugimoto, Sumiyoshi, Osaka 558-8585, Japan
\\
${}^{2}$Department of Physics, Kyoto University, Kyoto 606-8501, Japan
\\
${}^{3}$Kavli Institute for Theoretical Physics, Zhong Guan Cun East Street 55,
 Beijing 100080, P.R. China
}

\begin{abstract}
The stability of squashed Kaluza-Klein black holes is studied.
The squashed Kaluza-Klein black hole looks like a five dimensional 
black hole in the vicinity of horizon and and looks like a four
dimensional Minkowski spacetime with a circle at infinity.  In this sense, squashed Kaluza-Klein
black holes can be regarded as black holes in the Kaluza-Klein spacetimes.   
Using the symmetry of squashed Kaluza-Klein black holes, 
 $SU(2)\times U(1)\simeq U(2)$,  we obtain master equations for a part
 of the metric perturbations relevant to the stability.
  The analysis based on the master equations
  gives a strong evidence for the stability of squashed Kaluza-Klein black holes.
Hence, the squashed Kaluza-Klein black holes deserve to be taken seriously as 
realistic black holes in the Kaluza-Klein spacetime.  
\end{abstract}

\preprint{OCU-PHYS 285}
\preprint{AP-GR 51}
\preprint{KUNS-2111}
\preprint{CAS-KITPC/ITP-019}
\pacs{04.50.+h, 04.70.Bw}

\date{\today}
\maketitle

\section{introduction}

Recently, higher dimensional black holes have attracted much attention.
In particular, many exotic black holes in the asymptotically flat spacetime
are found~\cite{Myers:1986un,Emparan:2001wn,Pomeransky:2006bd,Elvang:2007rd,Iguchi:2007is,Izumi:2007qx,Elvang:2007hs}. 
From a realistic point of view, however, the extra dimensions need to be 
compactified to reconcile the higher dimensional gravity theory with 
our apparently four-dimensional world. 
The higher dimensional spacetimes with compact extra dimensions 
are called Kaluza-Klein spacetimes. 
The black holes should reside not in the asymptotically 
flat spacetimes but the asymptotically Kaluza-Klein spacetimes. 
We call these \lq Kaluza-Klein black holes\rq.
It would be important to study Kaluza-Klein black holes in the general dimensions. 
In this paper, we will consider five dimensional Kaluza-Klein black holes as a first step. 

It is well known that the simplest five-dimensional Kaluza-Klein black hole is 
the black string which is 
the direct product of four dimensional Schwarzschild black hole 
and a circle~\cite{Horowitz:1991cd}.
The topology of the horizon of black strings is $S^2 \times S^1$. 
The stability analysis of black strings has been done, and 
it was shown that black strings are stable when the horizon radius 
is larger than the scale of compact extra dimension~\cite{Gregory:1993vy}. 
Because of the stability, black strings are natural candidate of Kaluza-Klein
black holes.

Interestingly, another possibility has been recognized~\cite{Ishihara:2005dp}.
It is squashed Kaluza-Klein (SqKK) black holes that could also reside in 
the Kaluza-Klein spacetime. 
The topology of the horizon of SqKK black 
holes is $S^3$, while it looks like four dimensional black holes with
a circle as an internal space in the asymptotic region. 
SqKK black holes were originally derived as five-dimensional 
vacuum solutions in the context of Kaluza-Klein theory~\cite{Dobiasch:1981vh,Gibbons:1985ac}.
Recently, much effort has been devoted to reveal the properties of squashed 
Kaluza-Klein black holes~\cite{
Rasheed:1995zv,Larsen:1999pp,Cai:2006td,Wang:2006nw,Ishihara:2006iv,
Yazadjiev:2006iv,Brihaye:2006ws,Ida:2007vi,Ishihara:2006ig,Ishihara:2007ni,
Kurita:2007hu,Radu:2007te,Matsuno:2007ts,Yoo:2007mq,Chen:2007pu,Ishihara:2008re,Bizon:2006ue}.
Since the horizons of these black holes have the same nature as the five-dimensional black holes, 
Hawking radiation and quasi-normal modes from SqKK black holes 
would be different from those seen in four-dimensional black holes 
even at low energy~\cite{Ishihara:2007ni,Chen:2007pu,Ishihara:2008re}. 
That means that the extra dimension can 
be observed through these squashed black holes. 
These are distinct properties from black strings for which
we need to see the excitation of Kaluza-Klein modes to find the extra dimension. 
However, the stability of SqKK black holes is needed 
for these arguments to be meaningful. 

Related to the stability problem, Bizon et al~\cite{Bizon:2006ue} investigated 
the non-liner perturbation of Gross-Perry-Sorkin (GPS) monopole~\cite{Gross:1983hb,Sorkin:1983ns}
which is the zero mass limit of the SqKK black hole.
They showed GPS monopole is stable against small perturbations but
unstable against large perturbations
and collapses to a SqKK black hole.
This suggests that the SqKK black hole is a final
state of a gravitational collapse in the presence of GPS monopole.
Hence, SqKK black holes seem to be stable,
although the stability is not yet proved.
The purpose of this paper is to study the stability of SqKK black holes directly.

To analyze the stability, it is important to obtain a set of 
single ordinary differential equations of motion, the so-called master equations. 
To achieve this aim, we focus on the symmetry of SqKK black holes, 
$SU(2)\times U(1)\simeq U(2)$. Since SqKK black holes have the
same symmetry as five-dimensional Myers-Perry black holes with equal angular momenta, 
the analysis of field equations in the degenerate Myers-Perry spacetime~\cite{Murata:2007gv} 
can be applicable to SqKK black holes. 
By doing so, we show that metric perturbations which are supposed to be 
relevant to the stability can be described by master equations. 
Using the master equations, 
we prove the stability of SqKK black holes under these perturbations. 

The organization of this paper is as follows. 
In section~\ref{sec:SQBH}, we present the SqKK black
holes and discuss the symmetry of these black holes. 
In section~\ref{sec:Wigner}, the formalism to classify metric perturbations
is explained. Firstly, we introduce Wigner functions, which
are irreducible representation of $U(2)$. The tensor fields are expanded
in terms of these Wigner functions in invariant forms.
Using the classification based on the symmetry, we find infinite number of
master variables.  
In section~\ref{sec:stability}, we derive the master equations for
master variables. By analyzing these equations, we give a 
strong evidence of the stability of SqKK black holes. 
The final section is devoted to the discussion.

\section{Symmetry of squashed Kaluza-Klein black holes}\label{sec:SQBH}

In this paper, we concentrate on the static SqKK black holes in vacuum  
whose metric is given by
\begin{equation}
 ds^2 = -F(\rho)dt^2 + \frac{K(\rho)^2}{F(\rho)}d\rho^2 + \rho^2
  K(\rho)^2[(\sigma^1)^2+(\sigma^2)^2] +
  \frac{\rho_0(\rho_0+\rho_+)}{K(\rho)^2}(\sigma^3)^2\ .
\label{eq:sqbh}
\end{equation}
Here, the function $F(\rho)$ and $K(\rho)$ are given by 
\begin{equation}
 F(\rho)=1-\frac{\rho_+}{\rho}\ ,\quad K^2(\rho) =
  1+\frac{\rho_0}{\rho}\ ,
\end{equation}
where $\rho_+$ and $\rho_0$ are constant parameters.
The invariant forms $\sigma^a\,(a=1,2,3)$ of $SU(2)$ are given by
\begin{equation}
 \begin{split}
  \sigma^1 &= -\sin\psi d\theta + \cos\psi\sin\theta d\phi\ ,\\
  \sigma^2 &= \cos\psi d\theta + \sin\psi\sin\theta d\phi\ ,\\
  \sigma^3 &= d\psi + \cos\theta d\phi \ ,
 \end{split}
\end{equation}
which satisfy $d\sigma^a = 1/2 \epsilon^{abc} \sigma^b \wedge \sigma^c$, 
where $\epsilon^{abc}$ is the Levi-Civita symbol. 
The coordinate ranges are 
  $0\leq \theta \leq \pi $, $0\leq \phi \leq 2\pi$, $0\leq \psi \leq 4\pi$.

The angular part of the space, on which the metric \eqref{eq:sqbh} is spanned 
by $\sigma^a$, is topologically $S^3$. 
The horizon is located at $\rho=\rho_+$, and then its topology is $S^3$. 
In fact, the radius of $S^2$ is $\sqrt{\rho_+ (\rho_+ + \rho_0 )}$
 and the radius of the circle is $\sqrt{\rho_+ \rho_0}$. 
Hence, the geometry is a squashed three-sphere. 
The asymptotic form of metric at infinity becomes 
\begin{equation}
 ds^2 \sim -dt^2 + d\rho^2 + \rho^2 d\Omega_2^2 
+ \rho_0(\rho_0+\rho_+)(d\psi + \cos\theta d\phi)^2\ ,
\label{eq:asym_metric}
\end{equation}
where 
$d\Omega_2^2 = (\sigma^1)^2+(\sigma^2)^2 = d\theta^2+\sin^2\theta d\phi^2$ 
is the metric of $S^2$. From the metric~(\ref{eq:asym_metric}), 
we see the asymptotic geometry has the structure of $S^1$ fibered over $M^4$. 
Therefore, the extra dimension of spacetime~(\ref{eq:sqbh}) is compactified at
infinity, and the scale of compactification $\ell$ is given by
\begin{equation}
 \ell = \sqrt{\rho_0(\rho_0+\rho_+)}  \ .
\end{equation}
In this sense, the spacetime given by the metric~(\ref{eq:sqbh}), which has a 
squashed horizon, can be regarded as a kind of Kaluza-Klein black holes.  
Thus, the SqKK black hole has a $S^3$ horizon as a five-dimensional 
black hole and the asymptotic structure similar to that of a five-dimensional black string.  
It is well known that there exists Gregory-Laflamme instability~\cite{Gregory:1993vy} 
in the black string system. 
On the other hand, five-dimensional Schwarzschild black holes are stable~\cite{Ishibashi:2003ap,Konoplya:2007jv}. 
Therefore, it is interesting to study the stability of squashed black holes. 

Apparently, the metric~(\ref{eq:sqbh}) has the $SU(2)$ symmetry
generated by Killing vectors $\xi_\alpha \ , (\alpha=x,y,z)$:
\begin{equation}
\begin{split}
 \xi_x &= \cos\phi\partial_\theta +
 \frac{\sin\phi}{\sin\theta}\partial_\psi -
 \cot\theta\sin\phi\partial_\phi\ ,\\
 \xi_y &= -\sin\phi\partial_\theta +
 \frac{\cos\phi}{\sin\theta}\partial_\psi -
 \cot\theta\cos\phi\partial_\phi\ ,\\
 \xi_z &= \partial_\phi\ .
\end{split}
\end{equation}
The symmetry can be explicitly shown by using 
the relation $\mathcal{L}_{\xi_\alpha}\sigma^a=0$, 
where $\mathcal{L}_{\xi_\alpha}$ is a Lie derivative 
with respect to $\xi_\alpha$. 
The dual vectors to $\sigma^a$ are given by
\begin{equation}
\begin{split}
 \bm{e}_1 &= -\sin\psi \partial_\theta +
  \frac{\cos\psi}{\sin\theta}\partial_\phi - \cot\theta\cos\psi
 \partial_\psi\ ,\\
 \bm{e}_2 &= \cos\psi \partial_\theta +
  \frac{\sin\psi}{\sin\theta}\partial_\phi - \cot\theta\sin\psi
 \partial_\psi\ ,\\
 \bm{e}_3 &= \partial_\psi\ ,
\end{split}
\end{equation}
and, by definition, they satisfy $\sigma^a_i\bm{e}^i_b = \delta^a_b$.  
Let us define the two kind of angular momentum operators
\begin{equation}
  L_\alpha = i \xi_\alpha \ , \quad
  W_a  = i \bm{e}_a \ .
\end{equation}
where $\alpha,\beta,\cdots = x,y,z$ and $a,b,\cdots = 1,2,3$. 
They satisfy commutation relations
\begin{equation}
 [L_\alpha, L_\beta] = i \epsilon_{\alpha\beta\gamma} L_\gamma\ ,\quad
 [W_a, W_b] = -i\epsilon_{abc} W_c\ .
\end{equation}
They commute each other, $[L_\alpha, W_\alpha]$. 
From the metric (\ref{eq:sqbh}), we can also read off 
the additional $U(1)$ symmetry, which 
keeps the $S^2$ metric, $\sigma_1^2+\sigma_2^2$, invariant.
Thus, the spatial symmetry of SqKK black holes is $SU(2)\times U(1) \simeq U(2)$
\footnote{The metric~(\ref{eq:sqbh}) also has time translation 
symmetry generated by $\partial/\partial t$. 
},  
where $\bm{e}_3$ generates $U(1)$ and $\xi_\alpha\,(\alpha=x,y,z)$ generate $SU(2)$. 
As will be seen later, these symmetry yield the separability of equations 
for the metric perturbations. 

It is convenient to define the new invariant forms 
\begin{equation}
 \sigma^{\pm} = \frac{1}{2}(\sigma^1 \mp i \sigma^2)\ . 
\end{equation}
Here, we note that
\begin{equation}
	\mathcal{L}_{W_3} \sigma^{\pm} =  \pm  \sigma^{\pm} \ ,\quad 
	\mathcal{L}_{W_3} \sigma^3 = 0\ .
 \label{rule}
\end{equation}
The dual vectors to $\sigma^\pm$ are
\begin{equation}
 \bm{e}_{\pm} = \bm{e}_1 \pm i \bm{e}_2\ .
\end{equation}
By use of $\sigma^\pm$, the metric~(\ref{eq:sqbh}) can be rewritten as
\begin{equation} 
ds^2 = -F(\rho)dt^2 + \frac{K(\rho)^2}{F(\rho)}d\rho^2 + 4\rho^2
  K(\rho)^2\sigma^+ \sigma^- +
  \frac{\rho_0(\rho_0+\rho_+)}{K(\rho)^2}(\sigma^3)^2\ .
 \label{metric+-}
\end{equation}
%

\section{Classification of the metric perturbations based on the symmetry }\label{sec:Wigner}

Because the squashed black hole spacetime~(\ref{eq:sqbh}) has the 
$SU(2)\times U(1)$ symmetry, the metric perturbations can be expanded 
by the irreducible representation of $SU(2)\times U(1)$. 
We explain the formalism to obtain master equations 
for the metric perturbations~\cite{Murata:2007gv,Hu:1974hh}. 

Let us construct the representation of  $U(2) \simeq SU(2)\times U(1)$.
The eigenfunctions of $L^2\equiv L_\alpha^2 = W_a^2$ are degenerate, but
can be completely specified by eigenvalues of the operators $L_z$ and $W_3$. 
The eigenfunctions are called Wigner functions, which are defined by 
\begin{eqnarray}
 L^2 D^J_{KM} = J(J+1)D^J_{KM}\ ,  \quad
 L_z D^J_{KM} = M D^J_{KM}\ ,  \quad
 W_3 D^J_{KM} = K D^J_{KM}\ ,\label{eq:WigDef}
\end{eqnarray}
where $J,K,M$ 
satisfy $J\geq 0,\  |K|\leq J,\  |M|\leq J$.
From Eqs.~(\ref{eq:WigDef}), 
we see that $D^J_{KM}$ form the irreducible 
representation of $SU(2) \times U(1)$. 
The Wigner functions $D^J_{KM}(x^i)$ are functions defined on $S^3$, i.e., 
$x^i=\theta,\phi,\psi$, which satisfy the orthonormal relation
\begin{equation}
 \int^\pi_0 d\theta \int^{2\pi}_0 d\phi \int^{4\pi}_0 d\psi
 \sin\theta\, D^J_{KM}(x^i)D^{J'\,\ast}_{K'M'}{}(x^i) =
 \delta_{JJ'}\delta_{KK'}\delta_{MM'}\ .
\end{equation}

Now, we consider metric perturbations $g_{\mu\nu}+h_{\mu\nu}$, where $g_{\mu\nu}$ is the 
background metric \eqref{metric+-}. 
The tensor field $h_{\mu\nu}$ can be classified into three parts, 
$h_{AB},h_{Ai},h_{ij},~(A,B = t, \rho)$ which behave as scalars, vectors and a tensor under
the coordinate transformation of $\theta,\phi,\psi$. 
The scalars $h_{AB}$ can be expanded by the Winger functions as 
\begin{equation}
 h_{AB} = \sum_{K} h_{AB}^K(x^A) D_K(x^i)\ ,
\label{eq:hAB_dec}
\end{equation}
where we have omitted the indices $J,M$, because the metric perturbations
with different $J$ and $M$ are decoupled trivially in the perturbed equations. 

To decompose the vector part $h_{Ai}$, 
we construct vector harmonics as
\begin{equation}
\begin{split}
 &D_{i,K}^+ = \sigma^+_i D_{K-1}, \quad (|K-1|\leq J)\ ,\\
 &D_{i,K}^- = \sigma^-_i D_{K+1}, \quad (|K+1|\leq J)\ ,\\
 &D_{i,K}^3 = \sigma^3_i D_{K}, \qquad (|K|\leq J)\ .
\label{basis-v}
\end{split}
\end{equation}
One can check that 
\begin{equation}
 L^2 D_{i,K}^a = J(J+1)D_{i,K}^a\ ,\quad
 L_z D_{i,K}^a = M D_{i,K}^a\ ,\quad
 W_3 D_{i,K}^a = K D_{i,K}^a\ ,
\label{eq:vec_Wigner}
\end{equation}
where $a=\pm,3$ and operations are defined by Lie derivatives, that is, 
$W_a D_{i,K}^b \equiv \mathcal{L}_{W_a} D_{i,K}^b$ and 
$L_\alpha D_{i,K}^a \equiv \mathcal{L}_{L_\alpha} D_{i,K}^a$. 
In Eq.~(\ref{basis-v}), taking the relation (\ref{rule}) into account ,
we have shifted the index $K$ of Wigner functions 
so that $D_{i,K}^a$ have the same $U(1)$ charge $K$\cite{Murata:2007gv}. 
From Eqs.~(\ref{eq:vec_Wigner}),  we see that 
$D_{i,K}^a$ form the irreducible representation of $SU(2)\times U(1)$. 
Then, 
$h_{Ai}$ can be expanded as
\begin{equation}
 h_{Ai}(x^\mu) = \sum_K h_{Aa}^K(x^A)D_{i,K}^a(x^i)\ .
\label{eq:hAi_dec2}
\end{equation}

Similarly, the expansion of tensor part $h_{ij}$ can be carried out as 
\begin{equation}
 h_{ij}(x^\mu) 
= \sum_K h_{ab}^K(x^A)
  D_{ij,K}^{ab}(x^i)\ ,
\label{eq:hij_dec}
\end{equation}
where tensor harmonics $D_{ij,K}^{ab}$ are defined by
\begin{equation}
\begin{split}
&D_{ij,K}^{++} = \sigma^+_i \sigma^+_j D_{K-2} \quad(|K-2|\leq J)\ ,\\
&D_{ij,K}^{+-} = \sigma^+_i \sigma^-_j D_{K} \qquad(|K|\leq J)\ ,\\
&D_{ij,K}^{+3} = \sigma^+_i \sigma^3_j D_{K-1} \quad(|K-1|\leq J)\ ,\\
&D_{ij,K}^{--} = \sigma^-_i \sigma^-_j D_{K+2} \quad(|K+2|\leq J)\ ,\\
&D_{ij,K}^{-3} = \sigma^-_i \sigma^3_j D_{K+1} \quad(|K+1|\leq J)\ ,\\
&D_{ij,K}^{33} = \sigma^3_i \sigma^3_j D_{K} \qquad(|K|\leq J)\ . \\
\end{split}
\label{basis-t}
\end{equation}
We have shifted the eigenvalue $K$ of Wigner functions so that 
the tensor harmonics $D_{ij,K}^{ab}$ satisfy 
\begin{equation}
 L^2 D_{ij,K}^{ab} = J(J+1)D_{ij,K}^{ab}\ ,\quad
 L_z D_{ij,K}^{ab} = M D_{ij,K}^{ab}\ ,\quad
 W_3 D_{ij,K}^{ab} = K D_{ij,K}^{ab}\ . 
\label{eq:tensor_Wigner}
\end{equation}
Equations~(\ref{eq:tensor_Wigner}) mean that 
$D_{ij,K}^{ab}$ form the irreducible representation of $SU(2)\times U(1)$. 

Using the expansions (\ref{eq:hAB_dec}), (\ref{eq:hAi_dec2}) and (\ref{eq:hij_dec}) 
we can obtain a set of equations for expansion coefficient fields 
labelled by $J$, $M$, $K$. 
Because of $SU(2)\times U(1)$ symmetry no coupling appears between 
coefficients with different sets of indices $(J,M,K)$. 
 
Interestingly, without explicit calculation, we can 
reveal the structure of couplings between coefficients with the same $(J,M,K)$. 
First, since the index $K$ is shifted in the definition of 
vector and tensor harmonics, 
then the coefficients $h_{AB}^K$, $h_{Aa}^K$ and $h_{ab}^K$ exist 
for $K$ satisfying the inequality listed in the following table:

\begin{equation}
\begin{array}{|c|c|c|c|c|}
\hline
h_{++} & h_{A+},h_{+3} & h_{AB}, h_{A3} , h_{+-},h_{33}&h_{A-},h_{-3} &h_{--} \\ \hline
|K-2|\leq J &|K-1|\leq J    &|K|\leq J  &|K+1|\leq J   &|K+2|\leq J \\ \hline
\end{array}
\nonumber
\end{equation}
Therefore, for $J=0$ modes, we can classify the coefficients by possible $K$ as follows:

\hspace{2cm} $J=0;$
\begin{equation}
\begin{array}{|c|c|c|c|c|}
\hline
h_{++} & h_{A+},h_{+3} & h_{AB}, h_{A3} , h_{+-},h_{33}&h_{A-},h_{-3} &h_{--} \\ \hline
K=2    &               &                        &              &       \\ \hline
       &K=1            &                        &              &       \\ \hline
       &               &K=0                     &              &       \\ \hline
       &               &                        &K=-1          &       \\ \hline
       &               &                        &              &K=-2   \\ \hline
\end{array}
\nonumber
\end{equation}
Apparently, for $h_{++}$ and $h_{--}$, we can obtain the master equation for each variable, 
respectively.
For other sets of components, 
$(h_{A+},h_{+3})$, $(h_{AB}, h_{A3} , h_{+-},h_{33})$, $(h_{A-},h_{-3})$, 
they are coupled with each other in the same set. 
As we will see later, after fixing the gauge symmetry, 
we have the master equation for a single variable in each set. 
In total, there are five master equations, 
which matches the number of physical degrees of freedom of the gravitational 
perturbations. 

For $J=1$ modes, we can classify the coefficients as follows: 

\hspace{2cm} $J=1;$
\begin{equation}
\begin{array}{|c|c|c|c|c|}
   \hline
h_{++} & h_{A+},h_{+3} & h_{AB}, h_{A3} , h_{+-},h_{33}&h_{A-},h_{-3} &h_{--} \\ \hline
K=3    &               &                        &              &       \\ \hline
K=2    &K=2            &                        &              &       \\ \hline
K=1    &K=1            &K=1                     &              &       \\ \hline
       &K=0            &K=0                     &K=0           &       \\ \hline
       &               &K=-1                    &K=-1          &K=-1   \\ \hline
       &               &                        &K=-2          &K=-2   \\ \hline
       &               &                        &              &K=-3   \\ \hline
\end{array}
\nonumber
\end{equation}
We can see that $h_{++}$ in $(J=1,M,K=3)$ modes and 
$h_{--}$ in $(J=1,M,K=-3)$ modes are decoupled from other coefficients. 
It is easy to generalize this fact for arbitrary $J$, and 
we can also see that $h_{++}$ in $(J,M,K=J+2)$ modes and 
$h_{--}$ in $(J,M,K=-(J+2))$ modes are always decoupled. 
The perturbation equations for these modes can be reduced to the 
master equations for the single variables, respectively. 
%
\section{Stability Analysis of Squashed Kaluza-Klein Black Holes}
\label{sec:stability}

The 
gravitational perturbation 
equation in vacuum is 
\begin{eqnarray}
\delta R_{\mu \nu}
= \frac{1}{2}\big[
\nabla^\rho\nabla_\mu h_{\nu\rho} + \nabla^\rho\nabla_\nu h_{\mu\rho} -
  \nabla^2 h_{\mu\nu} - \nabla_\mu\nabla_\nu h
\big] 
= 0,
\label{eq:tensor_EOM}
\end{eqnarray}
where $\nabla_\mu$ denotes the covariant derivative with respect to the 
background metric $g_{\mu\nu}$ and $h= g^{\mu\nu} h_{\mu\nu}$. 
As is mentioned 
in the previous section, 
we can obtain master equations for variables in $(J=0,M=0,K=0,\pm1,\pm2)$ modes and 
$(J,M,K=\pm(J+2))$ modes. 
We derive these explicitly. 

\subsection{zero mode perturbations $(J=0)$}
In the case $J=0$, there are 
five physical degrees of freedom,
namely $K=\pm2,~ \pm1,~ 0$ modes. 
We treat these modes separately.

\subsubsection{$K=\pm 2$ modes}
\begin{figure}[htbp]
 \begin{center}
 \includegraphics[width=10cm,clip]{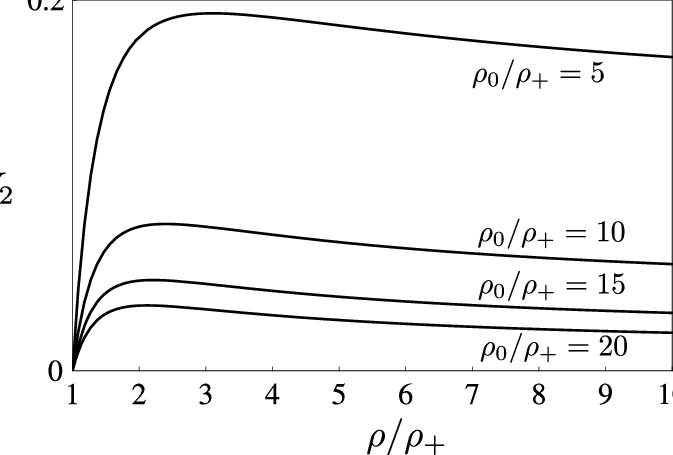}
 \end{center}
 \caption{
The effective potential $V_2$ for $K=\pm2$ mode.
}
 \label{figK2}
\end{figure}

In $K=\pm2$ modes, there exist two coefficients $h_{++}$ and $h_{--}$. 
We note that these are gauge invariant.
We consider only $h_{++}$ because $\bar h_{++}=h_{--}$, where bar denotes 
the complex conjugate. 
We set $h_{\mu\nu}$ as
\begin{equation}
 h_{\mu\nu}(x^\mu) dx^\mu dx^\nu = h_{++}(\rho)e^{-i\omega t}\sigma^+ \sigma^+\ .
\label{eq:h_K=2}
\end{equation}
Substituting Eq.~(\ref{eq:h_K=2}) into Eq.~(\ref{eq:tensor_EOM}), 
we get the equation of motion for $h_{++}$ as
\begin{eqnarray}
\delta R_{++} &= &
\frac{ {h_{++}}
       }{2\rho^2
     \rho_0{\left( \rho + \rho_0 \right) }^3
     \left( \rho_+ + \rho_0 \right) }
\Big[
4\rho^5 + 16\rho^4\rho_0 - 
       4\rho^3\left( \rho_+ - 5\rho_0 \right) \rho_0
\notag\\ && + 
       \rho_+\rho_0^3\left( \rho_+ + \rho_0 \right)  
+ 
       \rho\rho_0^2\left( 3\rho_+^2 + \rho_+\rho_0 + 2\rho_0^2 \right)
           + 4\rho^2\rho_0\left( \rho_+^2 + 3\rho_0^2 \right) 
\Big]
\notag\\ && - 
  \frac{ -2\rho^2 + 3\rho\rho_+ + \rho_+\rho_0 
    }{2\rho
     {\left( \rho + \rho_0 \right) }^2}  \frac{d h_{++}}{d \rho}
-
  \frac{\rho - \rho_+ 
     }{2
     \left( \rho + \rho_0 \right) } \frac{d^2 h_{++}}{d \rho^2}
-
  \frac{\rho}
   {2\left( \rho - \rho_+ \right) } \omega^2 {h_{++}} = 0.
\end{eqnarray}
In order to rewrite the equation in the Schr\"{o}dinger form, 
we introduce the new variable 
\begin{equation}
	\Phi_2(\rho) 
		\equiv \frac{1}{\rho^{1/4} (\rho + \rho_0)^{3/4}} h_{++}(\rho)\ ,
\end{equation}
and tortoise coordinate $\rho_*$ defined by
\begin{equation}
 \frac{d\rho_*}{d\rho} = \frac{K(\rho)}{F(\rho)}\ .
\label{eq:kame}
\end{equation}
Then, the final form of the equation becomes 
\begin{equation}
	-\frac{d^2}{d\rho_\ast^2}\Phi_2 + V_2(\rho) \Phi_2 = \omega^2 \Phi_2 \ ,
\label{eq:master_K2}
\end{equation}
where the potential $V_2(\rho)$ is defined by
\begin{equation}
\begin{split}
 V_2(\rho) 
		=& \frac{\rho-\rho_+}{16 \rho^3 \rho_0 (\rho_+ + \rho_0)(\rho +\rho_0)^3}
		\left[ 4 \rho_+ {\left( \rho_+ + \rho_0 \right) }^2 
		\left( 16 \rho_+^2 + 28 \rho_+ \rho_0 + 11 \rho_0^2 \right) \right. \\ 
	&+ \left( 320 \rho_+^4 + 960 \rho_+^3 \rho_0 + 996 \rho_+^2 \rho_0^2 + 
     391 \rho_+ \rho_0^3 + 35 \rho_0^4 \right)  
   \left( \rho - \rho_+ \right) \\
	&+ 8 \left( 80 \rho_+^3 + 182 \rho_+^2 \rho_0 + 127 \rho_+ \rho_0^2 + 
     25 \rho_0^3 \right)  {\left( \rho - \rho_+ \right) }^2  \\ 
	&+ 32 \left( 20 \rho_+^2 + 31 \rho_+ \rho_0 + 11 \rho_0^2 \right) 
		{\left( \rho - \rho_+ \right) }^3 \\ 
	&\left. + 64 \left( 5 \rho_+ + 4 \rho_0 \right)
		{\left( \rho - \rho_+ \right)}^4 + 64 {\left( \rho - \rho_+ \right) }^5
		\right]\ .
\label{potential2}
\end{split}
\end{equation}
From this expression, we can see $V_2>0$ in the region $\rho_+ <\rho <\infty$, 
explicitly. 
Typical profiles of the potential $V_2$ are plotted in Fig\ref{figK2}.

We consider that $\Phi$ is square integrable in the region $-\infty < \rho_*<\infty$. Then, $\omega^2$ is real.  
Multiplying both sides of Eq.~\eqref{eq:master_K2} by $\bar\Phi_2$ we have 
\begin{equation}
	-\bar\Phi_2 \frac{d^2}{d\rho_\ast^2}\Phi_2 + V_2(\rho) \bar\Phi_2\Phi_2 
		= \omega^2 \bar\Phi_2\Phi_2 .
\label{eq:K2}
\end{equation}
Adding eqs.\eqref{eq:K2} and its complex conjugate equation, 
and integrating it, we obtain 
\begin{equation}
 \int d\rho_\ast\left[
		  \left|\frac{d\Phi_2}{d\rho_\ast}\right|^2+V_2|\Phi_2|^2\right] 
	-\frac{1}{2}\left[ 
  \bar\Phi_2 {\frac{d}{d\rho_*}} \Phi_2+\Phi_2 {\frac{d}{d\rho_*}} \bar\Phi_2
					\right]^{\rho_\ast=\infty}_{\rho_\ast=-\infty}
 =\omega^2 \int d\rho_\ast |\Phi_2|^2 .
\label{eq:positive}
\end{equation}
Because the boundary term vanishes, the positivity of $V_2$ means $\omega^2>0$. 
Therefore, we have proved that the background 
metric is stable against the $K=\pm 2$ perturbations. 

\subsubsection{$K=\pm 1$ modes}
\begin{figure}[htbp]
 \begin{center}
 \includegraphics[width=10cm,clip]{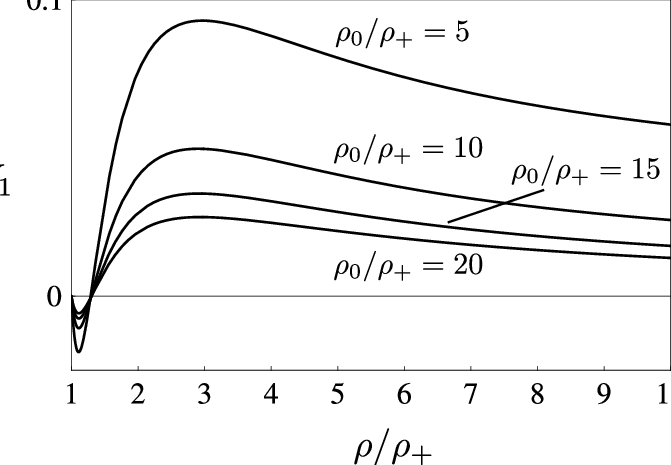}
 \end{center}
 \caption{
The effective potential $V_1$ for $K=\pm1$ mode.
}
 \label{figK1}
\end{figure}

Because of the relations $\bar h_{A+} =h_{A-}$ and $\bar h_{+3} =h_{-3}$, 
we consider only $h_{A+}$ and $h_{3+}$. 
We set $h_{\mu\nu}$ as
\begin{equation}
 h_{\mu\nu}dx^\mu dx^\nu = 2h_{A+}(\rho)\, e^{-i\omega t}\,dx^A\,\sigma^+ +
  2 h_{+3}(\rho)\, e^{-i\omega t}\,\sigma^+ \sigma^3\ .
\label{eq:h_K=1}
\end{equation}
There are three components in Eq.\eqref{eq:h_K=1}. 
The gauge transformations $h_{\mu \nu} \to h_{\mu \nu} + \nabla_{\mu}\xi_{\nu} + \nabla_{\nu}\xi_{\mu} $ 
for these variables are given by
\begin{eqnarray}
h_{t+} & \to & h_{t+} -i\omega \xi_+,
\\
h_{\rho+} & \to & h_{\rho+} - \frac{2\rho+\rho_0}{\rho(\rho+\rho_0)}\xi_+
+ \frac{d \xi_+}{d\rho},
\\
h_{3+} & \to & h_{3+} -\frac{i (\rho^2 +2 \rho \rho_0- \rho_+ \rho_0)}{(\rho + \rho_0)^2} \xi_+,
\end{eqnarray}
where we set $\xi_{\mu}dx^{\mu}$ as
\begin{eqnarray}
\xi_{\mu}dx^{\mu} &=&
\xi_{+}(\rho)e^{-i\omega t}\,\sigma^+.
\end{eqnarray}
So we can choose the gauge condition
\footnote{Note that 
we cannot choose this gauge condition
in the case of five dimensional Schwarzschild black hole limit.}
\begin{equation}
	h_{+3} =0\ ,
\label{eq:gauge_K1}
\end{equation}
which completely fixes the gauge freedom.  
Substituting Eqs.~(\ref{eq:h_K=1}) and (\ref{eq:gauge_K1}) into 
$\delta R_{A+} = 0$ and  $\delta R_{+3} = 0$, we obtain
\begin{eqnarray}
\delta R_{t+} & = & \frac{\rho^4 + \rho
        \left( 3\rho^2 - 2 \rho_+^2 \right)  \rho_0 + 
       \left( \rho -  \rho_+ \right) 
        \left( 3\rho +  \rho_+ \right)  \rho_0^2 + 
       \left( \rho -  \rho_+ \right)  \rho_0^3 
    }{2\rho^2 \rho_0
     {\left( \rho +  \rho_0 \right) }^2\left(  \rho_+ +  \rho_0 \right) 
     }h_{t +} 
\notag\\ && 
- \frac{i\left( \rho -  \rho_+ \right) \omega
    }{\rho
     \left( \rho +  \rho_0 \right) } h_{\rho +}
- 
  \frac{\left( \rho -  \rho_+ \right)  \rho_0
    }{2\rho
     {\left( \rho +  \rho_0 \right) }^2} \frac{d h_{t +}}{d \rho}
- 
  \frac{i\left( \rho -  \rho_+ \right) \omega
    }{2(\rho +  \rho_0)} \frac{d h_{\rho +}}{d \rho}
\notag\\ &&
- 
  \frac{\left( \rho -  \rho_+ \right) 
   }{2
     \left( \rho +  \rho_0 \right) }\frac{d^2 h_{t+}}{d \rho^2}
=0,
\\ 
\delta R_{\rho +} & = &
-\frac{i\omega
     \left( 2\rho +  \rho_0 \right)
     }{2\left( \rho -  \rho_+ \right) \left( \rho +  \rho_0 \right) }h_{t+}
\notag\\ &&
   -
\frac{h_{\rho +} }{2\rho
     \left( \rho -  \rho_+ \right)  \rho_0
     {\left( \rho +  \rho_0 \right) }^3\left(  \rho_+ +  \rho_0 \right) 
     }
\Big[
 -\rho^4
          \left( \rho -  \rho_+ \right)  
\notag\\ && + 
       \rho^2\left( -4\rho^2 + 6\rho \rho_+ - 2 \rho_+^2 + 
          \rho^3 \rho_+\omega^2 \right)  \rho_0 
\notag\\ &&+ 
       \left( {\left( -2\rho +  \rho_+ \right) }^2
           \left( -\rho +  \rho_+ \right)  + 
          \rho^4\left( \rho + 3 \rho_+ \right) \omega^2 \right) 
          \rho_0^2 
\notag\\ &&+ 3\rho^3\left( \rho +  \rho_+ \right) \omega^2
         \rho_0^3 + \rho^2\left( 3\rho +  \rho_+ \right) \omega^2
         \rho_0^4 + \rho^2\omega^2 \rho_0^5 
\Big]
\notag\\ &&+ \frac{i \rho\omega
     }{2(\rho -  \rho_+)}\frac{d h_{t +}}{d \rho}
=0,
\\ 
\delta R_{3 +} & = &
\frac{i \rho\omega 
     \left( \rho^2 + 2\rho \rho_0 -  \rho_+ \rho_0 \right) 
     }{2 \left( \rho -  \rho_+ \right) 
     {\left( \rho +  \rho_0 \right) }^2} h_{t+}
\notag\\ &&+ 
  \frac{2\rho^3 - 
       \rho^2\left(  \rho_+ - 5 \rho_0 \right)  + 6\rho \rho_0^2 - 
        \rho_+ \rho_0\left( 3 \rho_+ + 5 \rho_0 \right)  
    }{2
     {\left( \rho +  \rho_0 \right) }^4} h_{\rho +}
\notag\\ && + 
  \frac{\left( \rho -  \rho_+ \right) 
     \left( \rho^2 + 2\rho \rho_0 -  \rho_+  \rho_0 \right) 
    }{2
     {\left( \rho +  \rho_0 \right) }^3}\frac{d h_{\rho +}}{d \rho}
=0 \ .
\end{eqnarray}
Eliminating $h_{t+}$  from these equations, 
we get the master equation for $K=1$ mode. 
Defining a new variable
\begin{equation}
  \Phi_1(\rho) \equiv  \frac{4(\rho-\rho_+)
\left( \rho_+ \rho_0 - \rho \left( 2 \rho_0 + \rho \right)  \right)}
{\rho^{3/4}(\rho + \rho_0)^{9/4}} h_{\rho+}(\rho)\ , 
\end{equation}
we have the master equation in Shr\"odinger form:
\begin{equation}
 -\frac{d^2}{d\rho_*^2}\Phi_1 + V_1(\rho) \Phi_1 = \omega^2 \Phi_1\ .
\label{eq:master_K1}
\end{equation}
The potential $V_1$ reads
\begin{equation}
\begin{split}
V_1(\rho) =&  \frac{\rho-\rho_+}{
16 \rho_0 \left( \rho_+ + \rho_0 \right)  \rho^3 
  {\left( \rho_0 + \rho \right) }^3 
  {\left( \rho_+ \rho_0 - \rho 
       \left( 2 \rho_0 + \rho \right)  \right) }^2
}
\left[ 4 \rho_+^3 {\left( \rho_+ + \rho_0 \right) }^4 
   \left( 4 \rho_+^2 - 8 \rho_+ \rho_0 - 11 \rho_0^2 \right)\right.\\
&+ 
  \rho_+^2 {\left( \rho_+ + \rho_0 \right) }^3 
   \left( 144 \rho_+^3 + 48 \rho_+^2 \rho_0 - 68 \rho_+ \rho_0^2 + 
     31 \rho_0^3 \right)  \left( \rho - \rho_+ \right) \\ 
&+ 
  4 \rho_+ {\left( \rho_+ + \rho_0 \right) }^3 
   \left( 144 \rho_+^3 + 152 \rho_+^2 \rho_0 + 152 \rho_+ \rho_0^2 + 
     75 \rho_0^3 \right)  {\left( \rho - \rho_+ \right) }^2 \\
&+ 2 {\left( \rho_+ + \rho_0 \right) }^2 
   \left( 672 \rho_+^4 + 1520 \rho_+^3 \rho_0 + 1548 \rho_+^2 \rho_0^2 + 
     781 \rho_+ \rho_0^3 + 126 \rho_0^4 \right)  
   {\left( \rho - \rho_+ \right) }^3 \\ 
&+ 
  4 {\left( \rho_+ + \rho_0 \right) }^2 
   \left( 504 \rho_+^3 + 1032 \rho_+^2 \rho_0 + 757 \rho_+ \rho_0^2 + 
     191 \rho_0^3 \right)  {\left( \rho - \rho_+ \right) }^4 \\ 
&+ 
  \left( 2016 \rho_+^4 + 7072 \rho_+^3 \rho_0 + 9164 \rho_+^2 \rho_0^2 + 
     5211 \rho_+ \rho_0^3 + 1103 \rho_0^4 \right)  
   {\left( \rho - \rho_+ \right) }^5 \\ 
&+ 8 \left( 168 \rho_+^3 + 460 \rho_+^2 \rho_0 + 411 \rho_+ \rho_0^2 + 
     119 \rho_0^3 \right)  {\left( \rho - \rho_+ \right) }^6 \\ 
&+ 96 \left( 6 \rho_+^2 + 11 \rho_+ \rho_0 + 5 \rho_0^2 \right)  
   {\left( \rho - \rho_+ \right) }^7 \\ 
&+ \left. 16 \left( 9 \rho_+ + 8 \rho_0 \right)  
   {\left( \rho - \rho_+ \right) }^8 + 16 {\left( \rho - \rho_+ \right) }^9
\right]\ .
\end{split}
\label{eq:v1}
\end{equation}
Typical profiles of the potential $V_1$ are shown in Fig.\ref{figK1}.

From Fig.\ref{figK1}, we see that this potential $V_1$ contains a negative 
region. Hence, we hardly show the stability from this form of potential. 
However, we can overcome this difficulty by using a transformation 
of the coordinate. 
We introduce a new radial coordinate $y$ as
\begin{eqnarray}
\frac{d}{dy} = \frac{1}{\beta(\rho) }\frac{d}{d\rho_\ast} ,
\end{eqnarray}
where $\beta(\rho)$ is some real function and must be non-singular outside 
of the horizon, 
$\rho_+\leq \rho < \infty$.  
Then, the master equation becomes
\begin{eqnarray}
-\frac{d^2}{dy^2} \Phi_1
 -\frac{1}{\beta}\frac{d\beta}{dy}\frac{d}{dy}\Phi_1 
+\frac{V_1}{\beta^2} \Phi_1
=
\frac{\omega^2}{\beta^2} \Phi_1. \label{tildemastereqK1}
\end{eqnarray}
Multiply both sides of equation by $\bar \Phi_1$ we obtain
\begin{eqnarray}
-\bar \Phi_1\frac{d^2}{dy^2} \Phi_1
 -\frac{1}{\beta}\frac{d\beta}{dy}\bar \Phi_1\frac{d}{dy}\Phi_1 
+\frac{V_1}{\beta^2} \bar \Phi_1\Phi_1
=
\frac{\omega^2}{\beta^2} \bar \Phi_1\Phi_1. 
\label{barPhimasterPhi}
\end{eqnarray}
Adding eq.\eqref{barPhimasterPhi} and its complex conjugate equation, 
and integrating it, we obtain the equation
\begin{equation}
\begin{split}
\int dy 
\left[
\left| \frac{d \Phi_1}{dy}\right|^2 
+ 
\frac{\tilde{V}_1}{\beta^2} \left| \Phi_1 \right|^2 
\right]
&- \frac{1}{2}\left[\bar\Phi_1 \frac{d}{dy} \Phi_1 +\Phi_1\frac{d}{dy}\bar\Phi_1 
		+ \frac{1}{\beta}\frac{d\beta}{dy}
				|\Phi_1|^2\right]^{\rho_\ast=\infty}_{\rho_\ast=-\infty} \\
&=\omega^2 \int dy 
\frac{\left| \Phi_1 \right|^2}{\beta^2},
\end{split}
\label{omegapositivityK1}
\end{equation}
where
\begin{equation}
\tilde{V}_1 = V_1 + \frac{1}{2}\beta^2\frac{d}{dy}
\left(\frac{1}{\beta}\frac{d \beta}{dy}\right)\ .
\end{equation}
The boundary terms in (\ref{omegapositivityK1}) vanish because 
$\Phi_1$ is square-integrable.
Therefore, if
the deformed effective potential $\tilde{V}$ is positive everywhere,  
there are no $\omega^2 < 0$ mode.
Now, we choose $\beta$ as
\begin{eqnarray}
\beta^2 = \frac{15}{K(\rho)^2},
\end{eqnarray}
then the potential becomes
\begin{eqnarray}
\tilde{V}_1
&=&
 \frac{\rho-\rho_+}{
16 \rho_0 \left( \rho_+ + \rho_0 \right)  \rho^3 
  {\left( \rho_0 + \rho \right) }^3 
  {\left( \rho_+ \rho_0 - \rho 
       \left( 2 \rho_0 + \rho \right)  \right) }^2
}
\bigg[
16 \rho_+^3 {\left( \rho_+ - \rho_0 \right) }^2 
   {\left( \rho_+ + \rho_0 \right) }^4
\notag \\ && + 
  \rho_+^2 {\left( \rho_+ + \rho_0 \right) }^3 
   \left( 144 \rho_+^3 + 48 \rho_+^2 \rho_0 + 
     112 \rho_+ \rho_0^2 + 211 \rho_0^3 \right)  (\rho-\rho_+)
\notag \\ && + 
  4 \rho_+ {\left( \rho_+ + \rho_0 \right) }^3 
   \left( 144 \rho_+^3 + 152 \rho_+^2 \rho_0 + 
     152 \rho_+ \rho_0^2 + 75 \rho_0^3 \right)  (\rho-\rho_+)^2 
\notag \\ && + 
  2 {\left( \rho_+ + \rho_0 \right) }^2 
   \left( 672 \rho_+^4 + 1520 \rho_+^3 \rho_0 + 
     1248 \rho_+^2 \rho_0^2 + 361 \rho_+ \rho_0^3 + 6 \rho_0^4
     \right)  (\rho-\rho_+)^3 
\notag \\ && + 
  4 {\left( \rho_+ + \rho_0 \right) }^2 
   \left( 504 \rho_+^3 + 1032 \rho_+^2 \rho_0 + 
     532 \rho_+ \rho_0^2 + 11 \rho_0^3 \right)  (\rho-\rho_+)^4 
\notag \\ && +  
  \left( \rho_+ + \rho_0 \right)  
   \left( 2016 \rho_+^3 + 5056 \rho_+^2 \rho_0 + 
     3568 \rho_+ \rho_0^2 + 563 \rho_0^3 \right)  (\rho-\rho_+)^5 
\notag \\ && +  
  32 \left( \rho_+ + \rho_0 \right)  
   \left( 2 \rho_+ + \rho_0 \right)  
   \left( 21 \rho_+ + 26 \rho_0 \right)  (\rho-\rho_+)^6 
\notag \\ && + 
  96 \left( \rho_+ + \rho_0 \right)  
   \left( 6 \rho_+ + 5 \rho_0 \right)  (\rho-\rho_+)^7 
\notag \\ &&+ 
  16 \left( 9 \rho_+ + 8 \rho_0 \right)  (\rho-\rho_+)^8 + 
  16 (\rho-\rho_+)^9
~\bigg]\ .
\end{eqnarray}
We can see $\tilde{V}_1>0$ from above expression. 
Thus, we have proved the stability for $K=\pm 1$ modes.

\subsubsection{$K=0$ mode}
\begin{figure}[htbp]
 \begin{center}
 \includegraphics[width=10cm,clip]{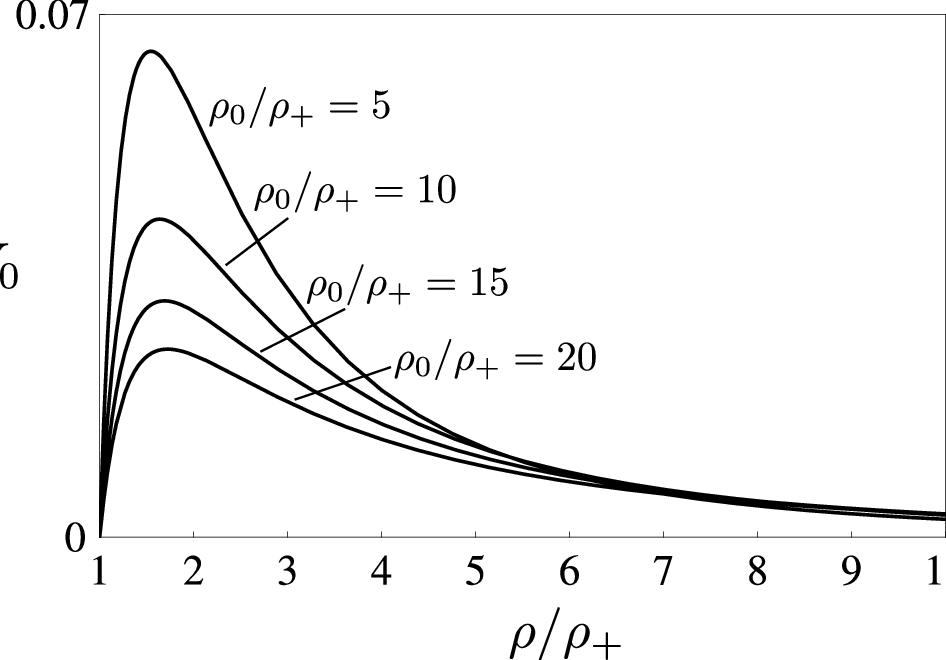}
 \end{center}
 \caption{
The effective potential $V_0$ for $K=0$ mode.
}
 \label{figK0}
\end{figure}

For $K=0$ mode, there are $h_{AB}, h_{A3}, h_{33}, h_{+-}$.
We set $h_{\mu\nu}$ as
\begin{multline}
 h_{\mu\nu}dx^\mu dx^\nu = h_{AB}(\rho)\,e^{-i\omega t} dx^A dx^B
+ 2h_{A3}(\rho)\,e^{-i\omega t} dx^A\sigma^3\\
+2h_{+-}(\rho)\,e^{-i\omega t} \sigma^+ \sigma^-
+ h_{33}(\rho)\,e^{-i\omega t} \sigma^3 \sigma^3\ .
\label{eq:h_K=0}
\end{multline}
The gauge transformations $h_{\mu \nu} \to h_{\mu \nu} + \nabla_{\mu}\xi_{\nu} + \nabla_{\nu}\xi_{\mu} $ 
for these variables are given by
\begin{eqnarray}
h_{tt} & \to & h_{tt}-2 i \omega \xi_t - \frac{\rho_+ (\rho - \rho_+)}{\rho^2(\rho + \rho_0)}\xi_{\rho},
\\
h_{t\rho} & \to & h_{t\rho} - \frac{\rho_+}{\rho (\rho-\rho+)}\xi_t -i\omega \xi_\rho + \frac{d \xi_t}{d \rho} ,
\\
h_{\rho\rho} & \to & h_{\rho\rho} + \frac{\rho_+ + \rho_0}{(\rho-\rho_+)(\rho+\rho_0)}\xi_\rho + 2 \frac{d\xi_\rho}{d\rho} ,
\\
h_{t3} & \to & h_{t3} -i \omega \xi_3,
\\
h_{\rho 3} & \to & h_{\rho 3} -\frac{\rho_0}{\rho (\rho + \rho_0)}\xi_3 + \frac{d\xi_3}{d\rho},
\\
h_{+-} & \to & h_{+-} +\frac{2(\rho -\rho_+)(2\rho + \rho_0)}{\rho + \rho_0}\xi_\rho,
\\
h_{33} & \to & h_{33} +\frac{(\rho - \rho_+)\rho_0^2 (\rho_+ + \rho_0)}{(\rho + \rho_0)^3}\xi_\rho,
\end{eqnarray}
where we set $\xi_{\mu}dx^{\mu}$ as
\begin{eqnarray}
\xi_{\mu}dx^{\mu} &=&
\xi_{A}(\rho)e^{-i\omega t}\,dx^A
+
\xi_{3}(\rho)\,e^{-i\omega t} \sigma^3.
\end{eqnarray}
So we can choose the gauge conditions
\footnote{Note that for static perturbation, 
we cannot choose this gauge condition.}
\begin{equation}
h_{+-} =0\ ,\quad
h_{tt} =0\ ,\quad
h_{t3} =0\ .
\label{eq:gauge_K0}
\end{equation}
Substituting Eq.~(\ref{eq:h_K=0}) and (\ref{eq:gauge_K0}) into
$\delta R_{AB} =0$, $\delta R_{33}=0$ and $\delta R_{+-}=0$, we have
\begin{eqnarray}
\delta R_{tt} & = & \frac{ \rho_+ \left( - \rho  +  \rho_+  \right)  \rho_0  + 
       2 \rho ^3{\left(  \rho  +  \rho_0  \right) }^2{\omega }^2
        }{4 \rho ^4 \rho_0 
     \left(  \rho  +  \rho_0  \right) \left(  \rho_+  +  \rho_0  \right) } 
h_{33}
- 
  \frac{i
     \left( 4 \rho  - 3 \rho_+  \right) \omega 
     }{ 2\rho 
     \left(  \rho  +  \rho_0  \right) } 
h_{t\rho}
\notag \\ &&+ 
  \frac{\left(  \rho  -  \rho_+  \right) 
     \left( - \rho_+ \left(  \rho_+  +  \rho_0  \right) 
           + 2 \rho ^2{\left(  \rho  +  \rho_0  \right) }^2
        {\omega }^2 \right) 
      }{4 \rho ^2
     {\left(  \rho  +  \rho_0  \right) }^3} 
h_{\rho \rho}
+ 
  \frac{\left(  \rho  -  \rho_+  \right)  \rho_+ 
      }{4 \rho ^3 \rho_0 
     \left(  \rho_+  +  \rho_0  \right) } 
\frac{d h_{33}}{d \rho}
\notag\\ &&- 
  \frac{i \left(  \rho  -  \rho_+  \right) \omega 
      }{ \rho  +  \rho_0 } 
\frac{d h_{t\rho}}{d \rho}
+ 
  \frac{\left(  \rho  -  \rho_+  \right)  \rho_+ 
     \left( - \rho  +  \rho_+  \right) 
      }{4 \rho ^2
     {\left(  \rho  +  \rho_0  \right) }^2}
\frac{d h_{\rho \rho}}{d \rho}
=0,
\\ 
\delta R_{t \rho} & = & -\frac{i \omega 
      }{4 \rho 
     \left(  \rho  -  \rho_+  \right)  \rho_0 } 
h_{33}
+ 
  \frac{i \left( - \rho  +  \rho_+  \right) 
     \left( 4 \rho  + 3 \rho_0  \right) \omega 
     }{4 \rho 
     {\left(  \rho  +  \rho_0  \right) }^2} 
h_{\rho \rho}
+ 
  \frac{i \left(  \rho  +  \rho_0  \right) 
     \omega  }{2 \rho  \rho_0 
     \left(  \rho_+  +  \rho_0  \right) }
\frac{d h_{33}}{d \rho}
=0,
\\ 
\delta R_{\rho \rho} & = &\frac{\left( - \rho  +  \rho_+  \right) 
     \left( 4 \rho ^2 - 2 \rho_+  \rho_0  + 
        \rho \left( -5 \rho_+  +  \rho_0  \right)  \right) 
      }{4 \rho ^3
     {\left(  \rho  -  \rho_+  \right) }^2
     \left(  \rho  +  \rho_0  \right) \left(  \rho_+  +  \rho_0  \right) } 
h_{33}
+ 
  \frac{i  \rho \left(  \rho_+  +  \rho_0  \right) 
     \omega  }{2{\left(
          \rho  -  \rho_+  \right) }^2\left(  \rho  +  \rho_0  \right) }
h_{t\rho}
\notag \\ && - 
  \frac{ \left(  \rho_+  +  \rho_0  \right) 
        \left( -4 \rho ^2 + 
          3 \rho \left(  \rho_+  -  \rho_0  \right)  + 2 \rho_+  \rho_0 
          \right)  + 2 \rho ^2
        {\left(  \rho  +  \rho_0  \right) }^3{\omega }^2 
        }{4 \rho 
     \left(  \rho  -  \rho_+  \right) {\left(  \rho  +  \rho_0  \right) }^3}
h_{\rho \rho}
\notag \\ &&
   + \frac{-2 \rho_+  \rho_0  + 
        \rho \left( - \rho_+  +  \rho_0  \right)  
      }{4 \rho ^2
     \left(  \rho  -  \rho_+  \right)  \rho_0 \left(  \rho_+  +  \rho_0  \right) }
\frac{d h_{33}}{d \rho}
  + \frac{i  \rho \omega 
      }{ \rho  -  \rho_+ } 
\frac{d h_{t\rho}}{d \rho}
\notag \\ &&+ 
  \frac{4 \rho ^2 - 3 \rho  \rho_+  + 3 \rho  \rho_0  - 
       2 \rho_+  \rho_0  
      }{4 \rho 
     {\left(  \rho  +  \rho_0  \right) }^2} 
\frac{d h_{\rho \rho}}{d \rho}
- 
  \frac{ \rho  +  \rho_0  
      }{2 \rho  \rho_0 
     \left(  \rho_+  +  \rho_0  \right) }
\frac{d^2 h_{33}}{d \rho^2}
=0,
\\
\delta R_{t 3} & = &\frac{-i \left(  \rho  -  \rho_+  \right) \omega 
      }{ \rho 
     \left(  \rho  +  \rho_0  \right) } 
h_{\rho 3}
- 
  \frac{i\left(  \rho  -  \rho_+  \right) 
     \omega  }{2( \rho  +  \rho_0) }
\frac{d h_{\rho 3}}{d \rho}
=0,
\\ 
\delta R_{\rho 3} & = &-\frac{ \rho {\omega }^2}
  {2( \rho  -  \rho_+ )}
h_{\rho 3}
=0,
\\ 
\delta R_{+-} & = &\frac{ 2 \rho 
          \left(  \rho  - 2 \rho_+  \right)  - 
         \left(  \rho  +  \rho_+  \right)  \rho_0   
       }{2 \rho ^2
     \left(  \rho  +  \rho_0  \right) \left(  \rho_+  +  \rho_0  \right) } 
h_{33}
+ 
  \frac{i  \rho \left( 2 \rho  +  \rho_0  \right) \omega 
      }{ \rho  +  \rho_0 } 
h_{t\rho}
\notag \\ &&+ 
  \frac{\left(  \rho  -  \rho_+  \right) 
     \left( 4 \rho ^2 +  \rho_0 \left( - \rho_+  + 3 \rho_0  \right)  + 
       2 \rho \left(  \rho_+  + 5 \rho_0  \right)  \right) 
      }{2
     {\left(  \rho  +  \rho_0  \right) }^3}
h_{\rho \rho}
\notag \\ && - 
  \frac{\left(  \rho  -  \rho_+  \right) 
     \left( 2 \rho  +  \rho_0  \right)  
     }{2 \rho  \rho_0 \left(  \rho_+  +  \rho_0  \right) } 
\frac{d h_{33}}{d \rho}
+ 
  \frac{{\left(  \rho  -  \rho_+  \right) }^2
     \left( 2 \rho  +  \rho_0  \right) 
      }{2
     {\left(  \rho  +  \rho_0  \right) }^2}
\frac{d h_{\rho \rho}}{d \rho}
=0,
\\ 
\delta R_{3 3} & = &\frac{ \left(  \rho  -  \rho_+  \right)  \rho_0 
          \left( 4 \rho  \rho_+  + 3 \rho  \rho_0  +  \rho_+  \rho_0  \right) - 
         2 \rho ^3{\left(  \rho  +  \rho_0  \right) }^3
          {\omega }^2  
       }{4 \rho ^2\left(  \rho  -  \rho_+  \right) 
     {\left(  \rho  +  \rho_0  \right) }^3}
h_{33}
\notag \\ && + 
  \frac{i  \rho  \rho_0 ^2
     \left(  \rho_+  +  \rho_0  \right) \omega 
      }{2{\left(  \rho  +  \rho_0  \right)
        }^3}
h_{t\rho}
 + \frac{3\left(  \rho  -  \rho_+  \right)  \rho_0 ^2
     {\left(  \rho_+  +  \rho_0  \right) }^2
    }{4
     {\left(  \rho  +  \rho_0  \right) }^5} 
h_{\rho \rho}
\notag \\ &&- 
  \frac{ 4 \rho ^2 - 2 \rho  \rho_+  +  \rho  \rho_0  +  \rho_+  \rho_0 
        }{4 \rho 
     {\left(  \rho  +  \rho_0  \right) }^2} 
\frac{d h_{33}}{d \rho}
+ 
  \frac{{\left(  \rho  -  \rho_+  \right) }^2 \rho_0 ^2
     \left(  \rho_+  +  \rho_0  \right) 
      }{4
     {\left(  \rho  +  \rho_0  \right) }^4} 
\frac{d h_{\rho \rho}}{d \rho}
\notag \\ &&- 
  \frac{ \rho  -  \rho_+  
      }{2
     \left(  \rho  +  \rho_0  \right) }\frac{d^2 h_{33}}{d \rho^2}
=0 \ .
\end{eqnarray}
Because of the gauge symmetry and constraint equations, 
there remains only one physical degree of freedom. 
In fact, introducing the new variable
\begin{equation}
\Phi_0(\rho) \equiv \frac{(\rho+\rho_0)^{5/4}(2\rho+\rho_0)}{\rho^{1/4}(4\rho +
 3\rho_0)}h_{33}(\rho)\ ,
\end{equation}
we get the the master equation 
\begin{equation}
 -\frac{d^2}{d\rho_*^2}\Phi_0 + V_0(\rho) \Phi_0 = \omega^2
  \Phi_0 ,
\end{equation}
where the potential $V_0$ is defined by
\begin{equation}
\begin{split}
V_0(\rho) =& \frac{\rho-\rho_+}{16\rho^3 (\rho+\rho_0)^3(4\rho+3\rho_0)^2}
\Big[
4 \rho_+ \left( 64 \rho_+^4 + 304 \rho_+^3 \rho_0 + 516 \rho_+^2 \rho_0^2 + 
     375 \rho_+ \rho_0^3 + 99 \rho_0^4 \right)  
\\ &+ 
  \left( 1024 \rho_+^4 + 3776 \rho_+^3 \rho_0 + 4656 \rho_+^2 \rho_0^2 + 
     2220 \rho_+ \rho_0^3 + 315 \rho_0^4 \right)  
   \left( \rho - \rho_+ \right)   \\ &+ 
  48 \left( 32 \rho_+^3 + 84 \rho_+^2 \rho_0 + 65 \rho_+ \rho_0^2 + 
     15 \rho_0^3 \right)  {\left( \rho - \rho_+ \right) }^2  \\ &+ 
  16 \left( 64 \rho_+^2 + 100 \rho_+ \rho_0 + 33 \rho_0^2 \right)  
   {\left( \rho - \rho_+ \right) }^3 + 
  128 \left( 2 \rho_+ + \rho_0 \right)
   {\left( \rho - \rho_+ \right) }^4 
\Big]\ .
\label{VK0}
\end{split}
\end{equation}
This expression explicitly shows $V_0>0$ outside the horizon. 
Then, we see the stability for $K=0$ mode.
Typical profiles of $V_0$ are shown in Fig.\ref{figK0}.

\subsection{$K =\pm(J + 2)$ mode perturbations}
As noted in the previous section, the highest modes, $h_{++}$ and $h_{--}$,
are always decoupled for arbitrary $J$. 
Since these are gauge invariant, 
it is straightforward to get the equation of motion for $h_{++}$ as
\begin{eqnarray}
\delta R_{++} &= &
\frac{ {h_{++}}
       }{2\rho^2
     \rho_0{\left( \rho + \rho_0 \right) }^3
     \left( \rho_+ + \rho_0 \right) }
\Big[
4\rho^5 + 16\rho^4\rho_0 - 
       4\rho^3\left( \rho_+ - 5\rho_0 \right) \rho_0
\notag\\ && + 
       \rho_+\rho_0^3\left( \rho_+ + \rho_0 \right)  
+ 
       \rho\rho_0^2\left( 3\rho_+^2 + \rho_+\rho_0 + 2\rho_0^2 \right)
           + 4\rho^2\rho_0\left( \rho_+^2 + 3\rho_0^2 \right) 
\notag\\ &&  + 
  J\rho{\left( \rho + \rho_0 \right) }^2
   \left( 4\rho^2 + 8\rho\rho_0 + 
     \rho_0\left( -\rho_+ + 3\rho_0 \right)  \right)
+
 J^2\rho{\left( \rho + \rho_0 \right) }^4 
\Big]
\notag\\ && 
- 
  \frac{ -2\rho^2 + 3\rho\rho_+ + \rho_+\rho_0 
    }{2\rho
     {\left( \rho + \rho_0 \right) }^2}  \frac{d h_{++}}{d \rho}
-
  \frac{\rho - \rho_+ 
     }{2
     \left( \rho + \rho_0 \right) } \frac{d^2 h_{++}}{d \rho^2}
-
  \frac{\rho}
   {2\left( \rho - \rho_+ \right) } \omega^2 {h_{++}}
=0.
\end{eqnarray}
Defining a new variable
\begin{equation}
\Phi_J(\rho) \equiv \frac{1}{\rho^{1/4} (\rho + \rho_0)^{3/4}} h_{++}(\rho)\ .
\end{equation}
we obtain the master equation
\begin{equation}
-\frac{d^2}{d\rho_\ast^2}\Phi_J + V_J(\rho) \Phi_J = \omega^2 \Phi_J \ ,
\label{eq:master_K=J+2}
\end{equation}
where the potential $V_J(\rho)$ is defined by
\begin{eqnarray}
V_J(\rho) &=&
\frac{\rho- \rho_+}{16\rho^3 \rho_0 (\rho_+ + \rho_0)(\rho + \rho_0)^3}
\bigg[
4 \rho_+ {\left( \rho_+ + \rho_0 \right) }^2 
   \left( 16 \rho_+^2 + 28 \rho_+ \rho_0 + 11 \rho_0^2 \right)  
\notag \\ &&+ 
  \left( \rho_+ + \rho_0 \right)  
   \left( 320 \rho_+^3 + 640 \rho_+^2 \rho_0 + 356 \rho_+ \rho_0^2 + 
     35 \rho_0^3 \right)  \left( \rho - \rho_+ \right)
\notag \\ &&  + 
  8 \left( \rho_+ + \rho_0 \right)  
   \left( 80 \rho_+^2 + 102 \rho_+ \rho_0 + 25 \rho_0^2 \right)  
   {\left( \rho - \rho_+ \right) }^2 
\notag \\ &&+ 
  32 \left( \rho_+ + \rho_0 \right)  
   \left( 20 \rho_+ + 11 \rho_0 \right)  
   {\left( \rho - \rho_+ \right) }^3 
\notag \\ &&+ 
  64 \left( 5 \rho_+ + 4 \rho_0 \right)  
   {\left( \rho - \rho_+ \right) }^4 + 
  64 {\left( \rho - \rho_+ \right) }^5
\notag \\ &&+J
\Big[
16 \rho_+ {\left( \rho_+ + \rho_0 \right) }^3 
   \left( 4 \rho_+ + 3 \rho_0 \right)  
\notag \\ &&+ 
  16 {\left( \rho_+ + \rho_0 \right) }^2 
   \left( 20 \rho_+^2 + 21 \rho_+ \rho_0 + 3 \rho_0^2 \right)  
   \left( \rho - \rho_+ \right)  
\notag \\ &&+ 
  16 \left( \rho_+ + \rho_0 \right)  
   \left( 40 \rho_+^2 + 53 \rho_+ \rho_0 + 14 \rho_0^2 \right)  
   {\left( \rho - \rho_+ \right) }^2 
\notag \\ &&+ 
  16 \left( \rho_+ + \rho_0 \right)  
   \left( 40 \rho_+ + 23 \rho_0 \right)  
   {\left( \rho - \rho_+ \right) }^3 
\notag \\ &&+ 
  64 \left( 5 \rho_+ + 4 \rho_0 \right)  
   {\left( \rho - \rho_+ \right) }^4 + 
  64 {\left( \rho - \rho_+ \right) }^5
\Big]
\notag \\ &&+J^2
\Big[
16 \rho_+ {\left( \rho_+ + \rho_0 \right) }^4 + 
  16 {\left( \rho_+ + \rho_0 \right) }^3 
   \left( 5 \rho_+ + \rho_0 \right)  \left( \rho - \rho_+ \right)  
\notag \\ &&+ 
  32 {\left( \rho_+ + \rho_0 \right) }^2 
   \left( 5 \rho_+ + 2 \rho_0 \right)  
   {\left( \rho - \rho_+ \right) }^2 
\notag \\ &&+ 
  32 \left( \rho_+ + \rho_0 \right)  
   \left( 5 \rho_+ + 3 \rho_0 \right)  
   {\left( \rho - \rho_+ \right) }^3 
\notag \\ &&+ 
  \left( 80 \rho_+ + 64 \rho_0 \right)  
   {\left( \rho - \rho_+ \right) }^4 + 
  16 {\left( \rho - \rho_+ \right) }^5 
\Big]
\bigg]\ .
\end{eqnarray}
Clearly, the potential $V_J$ is positive. 
Then, we confirm the stability against all $K=\pm(J+2)$ modes.

\section{summary and discussion}
We have studied the stability of squashed Kaluza-Klein (SqKK) black holes. 
By utilizing the symmetry $U(2)$ of the SqKK black holes, 
we have obtained the master equations for the metric perturbations 
labeled by $(J=0,M=0,K=0,\pm 1,\pm 2)$ and $(J,M,K=\pm (J+2) )$. 
We have proved the stability of SqKK black holes
for these perturbations. 
Strictly speaking, we have not shown the stability of SqKK
black holes completely, because we have analyzed the restricted modes. 
Empirically, the instability appear in the lower modes. 
For example, Gregory-Laflamme instability appears in a s-wave. 
Therefore, our result for $(J=0,M=0,K=0,\pm 1,\pm 2)$ modes gives 
a strong evidence for stability of the SqKK black holes.

Our stability analysis suggests that the SqKK black holes deserve 
to be taken seriously as 
realistic black holes in the Kaluza-Klein spacetime. 
Because of the stability, the SqKK black holes could be created in colliders 
or in the cosmic history. 
If so, we can observe the extra dimension through Hawking
radiation or quasi-normal modes~\cite{Ishihara:2007ni,Ishihara:2008re}. 
Namely, the SqKK black holes could be a window to the extra
dimension. 

There are several directions to be studied.
Our method can be applicable to other $U(2)$ symmetric spacetimes 
such as five-dimensional Myers-Perry black holes with equal angular
momenta\footnote{In the case of odd dimensions greater than five,
gravitational perturbations of Myers-Perry black holes with equal angular momenta
are discussed in \cite{Kunduri:2006qa}.}~\cite{Murata:2008yx}. 
The rotating SqKK black holes~\cite{Wang:2006nw}
has also the symmetry $U(2)$. It is known that
the rotation of black holes induces the superradiant instability for
massive scalar fields. 
Since Kaluza-Klein modes of gravitational perturbation 
are regarded as massive fields from the 4-dimensional point of view,
the rotating SqKK black holes may show
the superradiant instability. 
It is interesting to study if it occurs or not by using our formalism.
As an another direction, it is intriguing to study squashed black
holes in higher dimensions.

\section*{Acknowledgments}
We would like to thank Roman Konoplya for useful discussions. 
This work is supported by the Grant-in-Aid 
for Scientific Research No.19540305 and No. 18540262. 
K.M. is supported in part by JSPS Grant-in-Aid for Scientific Research,
No.193715 and also by the 21COE program ``Center for Diversity and
Universality in Physics,'' Kyoto University.
J.S. thanks the KITPC for hospitality during 
the period when a part of the work was carried out.

\newpage

\pagebreak

\begin{figure*}
{\large\bf Erratum}
\end{figure*}

In Sec.IV.A2 in the original paper, 
the discussion of the stability for $J=0, K=\pm 1$ modes contained an erroneous calculation. 
We correct it in this erratum.
First, we should note that the potential $V_1$ in Eq.~\eqref{eq:v1}
is non-negative 
in $\rho \ge \rho_+$ if 
$0 < \rho_0 \le \alpha_1$ with $\alpha_1 = 2 (\sqrt{15} -2)/11 \simeq 0.34$.
For the parameter region $\rho_0 > \alpha_1$, we use the $S$-deformation
method~\cite{Ishibashi:2003ap}, 
by which we can show the nonexistence of the unstable mode if the deformed potential
\begin{align}
{\tilde V}_1 = V_1 + F \sqrt{\frac{\rho}{\rho + \rho_0}}\frac{dS}{d\rho} - S^2,
\notag
\end{align}
is non-negative in $\rho \ge \rho_+$ for a continuous and bounded function $S$.
We find a function $S$  as
\begin{align}
S = \frac{F}{4 \rho_0 (\rho_++\rho_0)^2 \rho} \sqrt{1 + \frac{\rho_0}{\rho}} \left[
c_0 + c_1 X + c_2 X^2 + c_3 X^3
\right],
\label{sdeformationanalytic}
\end{align}
with $X = \rho_+ F/(\rho \rho_0 (\rho_++\rho_0))$ and
\begin{align}
c_0 &= -4 \rho_+^3+8 \rho_+^2 \rho_0+11 \rho_+ \rho_0^2,
\notag\\
c_1 &= 
2 \rho_+^5-16 \rho_+^4 \rho_0-26 \rho_+^3 \rho_0^2-3 \rho_+^2 \rho_0^3+3 \rho_+ \rho_0^4,
\notag\\
c_2 &=\frac{1}{3} \left(-4 \rho_+^7+52 \rho_+^6 \rho_0-100 \rho_+^5 \rho_0^2-468 \rho_+^4 \rho_0^3-333
   \rho_+^3 \rho_0^4+30 \rho_+^2 \rho_0^5+42 \rho_+ \rho_0^6\right),
\notag\\
c_3 &=\frac{1}{12} \big(11 \rho_+^9-204 \rho_+^8 \rho_0+954 \rho_+^7 \rho_0^2+740 \rho_+^6
   \rho_0^3-6957 \rho_+^5 \rho_0^4-11736 \rho_+^4 \rho_0^5-5160 \rho_+^3 \rho_0^6
\notag\\ & \quad +468 \rho_+^2 \rho_0^7+372
   \rho_+ \rho_0^8\big).
\notag
\end{align}
These coefficients can be obtained by imposing ${\tilde V}_1 = {\cal O}((\rho -\rho_+)^5)$ near the horizon.
This function $S$ gives non-negative $\tilde{V}_1$ in $\rho \ge \rho_+$
for the parameter region
$ \alpha_1 < \rho_0/\rho_+ < \alpha_2 $,
where $\alpha_2 \simeq 4.72$ is the largest root of a polynomial
\begin{align}
&-2700 x^{10}-5520 x^9+46560 x^8+153600 x^7+153901 x^6+29828
   x^5-28699 x^4-5056 x^3
\notag\\ & +3461 x^2
-460 x+19,
\notag
\end{align}
and it
corresponds to $ r_\infty/\rho_+ \simeq 10.4$.

\begin{figure}[h]
\includegraphics[width=10cm]{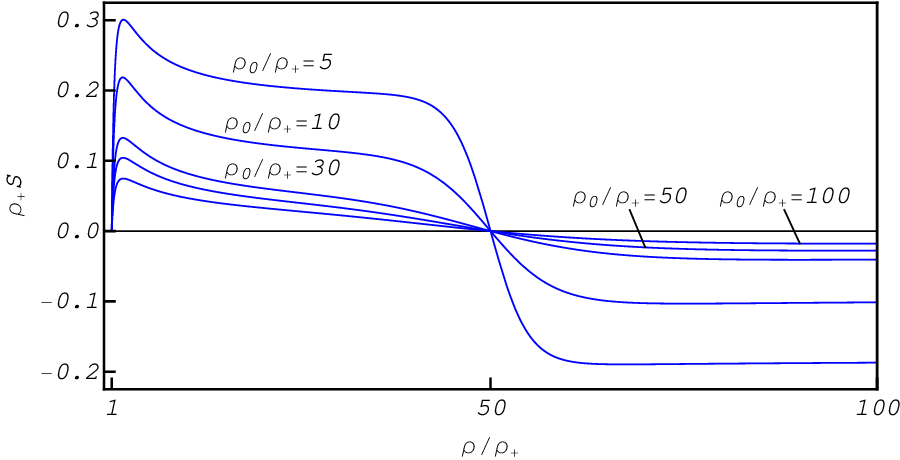}
\caption{\label{fig} 
The deformation functions $S$ for various $\rho_0$ with the boundary condition $S|_{\rho = 50 \rho_+ } = 0$.
 }
\end{figure}

When $\rho_0$ is large, it is hard to find an appropriate function $S$ analytically,
in fact, we cannot find it by adding the higher order terms of $X$ to the rhs in Eq.\eqref{sdeformationanalytic}.
However, we can still numerically find the deformation function by 
solving the equation ${\tilde V}_1 = 0$ as shown in~\cite{Kimura:2017uor}.
In Fig.\ref{fig}, we plot the numerical solutions for various $\rho_0$. 
This shows the nonexistence of unstable mode for large $\rho_0$.
While there is already a numerical proof of the stability based on the quasinormal mode~\cite{Ishihara:2008re}, this is a complimentary result that also supports the stability by a different way.

We thank Ryusuke Nishikawa for pointing out this problem
and useful discussion.

\end{document}